\documentclass[useAMS,usenatbib]{mn2e}
\usepackage{aas_macros}
\usepackage{amsmath}
\usepackage[varg]{txfonts}
\usepackage{graphicx}
\usepackage{epsfig}
\usepackage{amssymb}

\title{Photometry and H$\alpha$ studies of a Low Mass Ratio Overcontact binary ASAS J082243+1927.0}
\author[K. Sriram, D. Shanti Priya \& P. Vivekananda Rao]{K. Sriram$^{1}$\thanks{E-mail:
astrosriram@yahoo.co.in.}, D. Shanti Priya$^{1}$ \& P. Vivekananda Rao$^{1}$\\
$^{1}$Department of Astronomy, Osmania University, Hyderabad 500 007, India}

\begin{document}

\pagerange{\pageref{firstpage}--\pageref{lastpage}} \pubyear{2014}

\maketitle

\label{firstpage}

\begin{abstract}
Both high precision CCD photometric and H$\alpha$ line studies are presented for an overcontact binary ASAS J082243+1927.0. The light curve exhibits a total eclipse at secondary minima along with an {\bf O'Connell} effect. The light curve was modeled using the Wilson Devinney code and the best solution provides the mass ratio q $\sim$ 0.106 and fill-out factor f $\sim$ 72\% . These parameters indicate that the system is a low mass ratio overcontact binary with a high degree of geometrical contact. The H$\alpha$ line equivalent width varied at different phases and it is found that the line is possibly filled-in at secondary minima. From a small sample of overcontact binaries, we found a correlation between the orbital period and H$\alpha$ line equivalent width of the primary component. Based on a sample of high filling factor and low mass ratio contact binaries, a mass ratio cut-off is observed at q$_{critical}$ = 0.085 in mass-ratio -- period plane. It was observed that below q$_{critical}$ $<$ 0.085, period decreases with an increase in q and above it, period increases as the mass ratio increases. Interestingly, the observed mass ratio cut-off value lies close to the critical mass ratio range as predicted in the literature. The observational evidence of the cut-off of the mass ratio and its variation with orbital period {\bf are} discussed in terms of mass transfer and angular momentum loss. Based on the results, we suggest that, ASAS J082243+1927.0 is at the verge of merger, eventually forming a fast rotating star.

\end{abstract}


\section{Introduction}

W Ursae Majoris (W UMa) variables are eclipsing overcontact binaries with orbital periods ranging from 0.2 -- 1.0 day. These systems consist of main sequence stars with spectral A-K type sharing a common convective envelope due to filled Roche lobes. In some cases these binaries host O or B spectral type component surrounded with a common radiative envelope, whose true physical understanding is still lacking. In low mass overcontact binaries, it has been argued that the secondary component is oversized with respect to its expected ZAMS radius and at an advanced evolutionary stage (Stepien 2006a). The role of common envelope is to distribute the energy uniformly over the surface of the stars (Lucy 1968), having similar brightness with a few percent difference exhibiting chromospheric activity (Vilhu \& Walter 1987). The overcontact binaries are important astrophysical sources as they help to understand the underlying mechanism of the merging process (eg. V 1309 Sco; Tylenda et al. 2011), stellar dynamo process (eg. Qian et al. 2005), contributing in understanding the galactic structure because of their high number density (1/500 MS stars; Rucinski 2002), binary evolution theories (eg. Yakut \& Eggelton 2005) and also serve as distance estimators (Rucinski \& Duerbeck 1997). All the overcontact binaries are classified in three broad categories, A-type, W-type (Binnendijk 1970) and B-type (Csizmadia \& Klagyivik 2004). In the A-type, the less massive component eclipses the massive one causing the primary minimum and an opposite scenario is observed in case of the W-type. In general, A-types often have low mass ratio (q $<$ 0.3), relatively long orbital periods (P $>$ 0.3 days), whereas W-types have mass ratios, q $>$ 0.3 and short orbital periods (P $<$ 0.3 days). It {\bf had been suggested earlier that A-types are in an} advanced evolutionary stage compared to the W-types (eg. Hilditch 1989) but later was overruled as A-types have more mass and angular momentum (Gazeas \& Niarchos 2006). But a number of overcontact binaries known to harbor a third component causing the sinking of angular momentum and hence the discrepancy in the evolutionary status could be resolved by constraining their age. The high temperature difference $>$ 1000 K between the components in overcontact binaries forms the basis for B-type classification (Csizmadia \& Klagyivik 2004) and {\bf systems in this class} are also known as poor thermal overcontact systems (Rucinski \& Duerbeck 1997). 

Many of the close binaries of several types and overcontact binaries light curves exhibit asymmetry in the brightness of maximum light, known as {\bf O'Connell} effect ({\bf O'Connell} 1951; Milone 1969; Davidge \& Milone 1984) and is often associated with a dark spot on the primary component. The strong evidence for the presence of the spot comes from the study of H$\alpha$ line in overcontact binary systems. The first detailed study was performed by Barden (1985) on four W UMa systems showing that the H$\alpha$ line is a strong signature of the magnetic-associated activity in these systems. The study of H$\alpha$ line is also important as the magnetic field plays a key role in the evolution of overcontact binaries via the magnetic braking process (Stepien 1995). The presence of spots are related to the chromospheric activity causing the filling of the H$\alpha$ line and varying the equivalent width along with the orbital period (Kaszas et al. 1998). Moreover due to this activity overcontact binary systems are also good X-ray emitters (McGale et al. 1996; Stepien 2001; Chen et al. 2006) and the related X-ray emission is connected to the stellar dynamo activity arising from the synchronous fast rotating convective common envelope (Gondoin 2004). An illustrative study of VW Cep (Kaszas et al. 1998), AE Phe and YY Eri (Maceroni et al. 1994; Vilhu and Maceroni 2007) clearly suggest that the activity is related to primary/massive component (as it has deep convective zones) which is in agreement with the theoretical studies (Rucinski 1992, 1994).

Deep (f $\ge$ 50 \%) low mass ratio (q $\le$ 0.25) overcontact binaries (DLMR) are considered to be important sources and are possible progenitors for FK Com-type and blue stragglers (Qian et al. 2006). {\bf Although} a different naming/classification {\bf was} adopted, most of them are A-type overcontact binaries. They have a period domain ranging from 0$^{d}$.2709 (J13031-0101.9) to 0$^{d}$.8664 (KN Per). Qian et al. (2006) found that few of the systems undergo secular period decrease. The coupled action of angular momentum loss (AML) and thermal relaxation oscillation (TRO; Lucy 1976; Flannery 1976; Robertson \& Eggleton 1977) in the overcontact binary, leads to increase in the lifetime of overcontact phase. At this stage the binary can meet Hut's criteria i.e. J$_{rot}$ $>$ 1/3 J$_{orb}$ (Hut 1980) or can encounter dynamical instability (Rasio \& Shaprio 1995) which results in merging of the components. Such mergers are rare but V1309 Sco can be considered as a prototype for such events. Tylenda et al. (2011) concluded that V1309 Sco was a cool overcontact binary system and instabilities caused the secular period decrease of about 24.5 min over a duration of six years. Based on the formation models of the cool overcontact binaries (Stepien 2004, 2006a,b, 2009), Stepien (2012) concluded that the loss of mass and angular momentum through magnetic winds played a crucial role in the merging process.   

The variability of ASAS J082243+1927.0 (V1) at $\alpha$$_{J2000}$=08$^{h}$ 22$^{m}$ 43$^{s}$.0 and $\delta$$_{J2000}$=+19$^{\circ}$ 26$^{`}$ 58$^{``}$ was discovered in ASAS (ASAS J082243+1927.0; Pojamanski 2002). Later light curves of this system were reported by Gettel et al. (2006) with star ID 10100383, Pepper et al. (2008; Id KP400793) and Terrell et al. (2012). Due to the small orbital period 0.$^{d}$28 and shape of the light curve, it was classified as an EW type variable. The colours of the variable are B-V = 0.57 (Terrell et al. 2012), J-H = 0.31 and H-K = 0.33 (Gettel et al. 2006). Due to the large scatter in the earlier reported light curves, the nature of the eclipses and presence of any O'Connell effect were not clearly visible. V1 exhibits total eclipses and hence mass ratio can be accurately estimated for such systems (Terrell \& Wilson 2005). Moreover, study of the H$\alpha$ line among overcontact binaries has been rare when compared to photometry. Keeping this objective in view, we performed V band CCD photometry and spectroscopy concentrating on the variation of H$\alpha$ line at various phases for this variable.

\section{Observations, Data Reduction and Analysis }
The V band photometric observations were made on January 19, 20 and 22, 2013 with the 2.0-m telescope of the IUCAA-Girawali Observatory (IGO). IUCAA Faint Object Spectrograph and Camera (IFOSC) was used which employes a 2K $\times$ 2K, thinned, back-illuminated CCD with a pixel size of 13.5$\mu$m, gain 1.5e$^{-}$ / ADU and read out noise 4e$^{-}$ . The CCD provides a 10.5$^{'}$ $\times$ 10.5$^{'}$ image with a plate scale of 0.3 arcsec pixel$^{-1}$. The observations were performed in the air mass range 1.01 -- 1.80. The images were acquired in Bessell's V filter with an integration time of 10 s. The extinction corrections were not {\bf applied} as the comparison is close to the variable and of similar brightness (see Figure 1) and transformation to the standard system was not done. The Figure 1 shows the location of variable (V1), comparison (TYC 1386-1630-1) and check (TYC 1386-121-1) stars. The V magnitudes and colour indexes of variable, comparison and check stars are shown in Table 1. These stars were selected as they are nearby and of similar brightness to V1. The magnitude difference between variable and comparison stars\footnote{The data will be made available from the authors on request.} along with check and comparison stars observed on 19 January 2013 are shown in Figure 2. The magnitude difference between check and comparison stars was found to be constant $\sim$ 0.02 $\pm$ 0.002.

Spectroscopic observations were performed on December 5--6, 2013, with the 2.3-m Vainu Bappu telescope (VBT) of the Vainu Bappu Observatory (VBO) equipped with the Optomechanics Research (OMR) spectrograph along with a detector of 1K $\times$ 1K CCD. The obtained spectra cover a range of 3000\AA\  centered around H$\alpha$ line with a 600 lines/mm dispersion resulting in a resolution of 2.6 \AA\ /pixel. The exposure time was 35--40 min for both the variable V1 and a spectrophotometric standard (BD+08 2015). The {\it FeNe} arc lamp was observed for wavelength calibrations. 
We used different packages made available in {\it IRAF}\footnote{IRAF is distributed by the National Optical Astronomy Observatory, which is operated by the Association of Universities for Research in Astronomy (AURA) under cooperative agreement with the National Science Foundation.} to reduce the data.  In both photometric and spectroscopic reduction, bias and flat-field correction were made and later {\it APPhot} and {\it ONEDSPEC} packages were used to measure the magnitudes and to obtain the spectra. Later the spectra were normalized for further studies.

\subsection{Period analysis and determination of ephemeris}
We determined the period of the variable using {\it Period04} package (Lenz \& Breger 2005) and times of minima using the Kwee \& van Woerden’s method (1956). The ephemeris determined for the variable is Min${_I}$ = 2456312.2997(89) + 0$^{d}$.28000(2)E. The obtained period is similar to the one reported in previous studies.  

\subsection{Photometric solution using Wilson Devinney method}
The V band light curve covering all the phases with 418 data points is shown in Figure 3. Visual inspection of the light curve clearly shows the total eclipse at secondary minimum along with an O'Connell effect notable at phase 0.25. Based on the colour indices of the variable, i.e., B-V=0.57 and J-H=0.31, the primary component temperature was fixed at 5960 K. The photometric solutions were obtained using the latest version of Wilson-Devinney (W-D) code v2013 with an option of non-linear limb darkening via a square root law (Wilson \& Devinney 1971; Wilson 1979; Wilson 1990; Van Hamme \& Wilson 2007; Wilson 2008; Wilson \& Van Hamme 2013). The new code has 60 parameters among which 50 are {\bf adjustable}. In this version, a high precision star spot algorithm is incorporated which allows {\bf one} to develop time varying spots by adjusting times of onset, size, and disappearance which affects the light curve{\bf ;} the new code also permits the spot to drift, independent of star rotation (Wilson 2012). The distance of binaries can be estimated via Direct Distance Estimation (DDE) process using a control integer IFCGS=1 in the code, but to adjust this parameter, the code requires well calibrated photometry in at least two passbands (Wilson et al. 2010). We adopted the following {\bf model and} method (eg. Ravi, Sriram \& Vivekananda Rao 2012; Shanti, Sriram \& Vivekananda Rao 2013). A convective outer envelope was assumed for both components; gravity darkening co-efficients $g_1 = g_2$ = 0.32 (Lucy 1967) and bolometric albedos  $A_1 = A_2$ = 0.5 (Rucinski 1969) were taken as fixed parameters. The value of the limb darkening coefficients of components $x_1, x_2$ were fixed at 0.64 (Van Hamme 1993) for V band. The adjustable parameters were the following: temperature of secondary component $T_{2}$, orbital inclination ({\it i}), the dimensionless potentials of the primary component $\Omega_1$= $\Omega_2$ and the bandpass luminosity of the primary star $L_1$. The overcontact configuration i.e. mode 3 option of W-D method was used to determine the parameters after overruling the mode 2 option.\\

As no spectroscopic mass ratio was determined for this variable, the grid search method was adopted. The total eclipse at secondary minimum suggests that it is a low mass ratio system (0.1 $<$ q $<$ 0.2) (Terrell \& Wilson 2005; Wilson 2006). We searched the parameter space of q in the range of 0.02 $<$ q $<$ 10.0 along with other adjustable parameters. During the computations, the solutions gradually converged from detached (mode 2) to overcontact (mode 3) configuration. The resulting sum of the weighted square deviations, $\sum (\omega_(o-c))^2$ over the selected range of mass ratio q were noted and a minima at q = 0.121 was observed. Later, it was made as an adjustable parameter along with others in the differential correction routine. During the computations, high residuals were observed at phase 0.25 and hence a solution with a dark spot over primary / secondary component and group of spots were tested to model the observed O'Connell effect. First computation was {\bf done} with an {\bf application of a dark spot} over {\bf the} primary, which resulted in lowering the residuals (Dark 1 in Table 2; Fig. 3, top panel). A relatively dark and large spot over {\bf the} secondary, resulted in a good fit (Dark 2 in Table 2). We also attempted a solution by varying the spot longitude over {\bf the} primary (IFSMV1=1) and drift of the spot was made independent {\bf of} the star rotation (eg. F1$_{spot}$=0.80) but the overall solution remained almost the same except that the longitude and latitude of the spot varied slightly for the fixed size of the spot. If F1$_{spot}$ is varied below 0.8 then the solution was found to be diverging. We also found that {\bf the} secondary temperature is higher than the primary's although an occultation is observed at secondary minimum (Table 2).  

Based on the solutions, we found that the best value of mass ratio to be q $\sim$ 0.106. To check the consistency of the solution, the temperatures and inclination were varied by 5--10\% and no significant difference in the respective solutions were found. The Table 2 shows the photometric elements obtained from the best solutions and the corresponding theoretical fits are shown in Figure 3. The fill-out factor or degree of overcontact parameter, f = $\frac {\Omega_{in} - \Omega_{1,2}} {\Omega_{in} - \Omega_{out}}$ was derived for all the solutions and were found to be in the range of $\sim$ 61--72 \%. The obtained values of q and f reveals that the variable is a low mass ratio overcontact binary with a high degree of geometrical contact.  \\
 
\subsection{H$\alpha$ 6563 \AA\  emission study}
The H$\alpha$ line was observed at phases: 0.15, 0.33, 0.51 and 0.81 for the variable and its mean equivalent width (EW) was found to be 1.60 $\pm$ 0.13 \AA\ . Figure 4 shows the variation of the H$\alpha$ profiles with respect to the spectrophotometric standard. It is evident that at phase 0.51, the variable H$\alpha$ line is possibly filled-in with respect to the standard star at a significance of about 2.7$\sigma$ whereas at other phases, {\bf the} filled-in feature is relatively weak. The {\bf fill-in}, especially at phase 0.5 is possibly caused by chromospheric emission. This is also supported by the observed O'Connell effect exhibited in the light curve due to {\bf the} presence of a dark spot on the primary or secondary. As continuous spectroscopic observations are lacking, determining the differences in the profiles arising from individual component and obtaining the H$\alpha$ EW for each component were not possible. However at phase 0.5, when the secondary is being eclipsed, we argue that the EW$_{p}$$^{i}$ = 1.09\AA\ of H$\alpha$ line is most probably arising from the primary component. The intrinsic equivalent width of the line at phase 0.5 was calculated using the relation \\
EW$_{p}$$^{i}$=a $\times$ EW$_{p}$ and EW$_{s}$$^{i}$=a $\times$ EW$_{s}$, where a = 1 + q$^{0.92}$ ($T_{s}/T_{p}$)$^{4}$ (Webbink 2003), EW$_{p}$, EW$_{s}$ are measured equivalent widths and EW$_{p}$$^{i}$, EW$_{p}$$^{i}$ are intrinsic equivalent widths. A small sample of periods and EWs for overcontact binaries viz. AE Phe, YY Eri (Vilhu and Maceroni 2007), VW Cep (Kaszas et al. 1998) suggests a correlation between the two parameters (Fig. 4). Since the conclusion of presence of the H$\alpha$ filled-in was limited to visual inspection and hence a continues spectroscopic monitoring of the variable is important to unveil the nature of the H$\alpha$ line.   \\

\section{Discussion and Conclusion}
The V band light curve shows a total eclipse at secondary minimum and exhibits {\bf the O'Connell} effect. These features were not clear in the previous published light curves. However the variable O'Connell effect is not unusual in overcontact binaries. Based on the derived solutions, perhaps the variable can be classified as a W-type W UMa system, {\bf as the} secondary temperature was found to be relatively higher (Table 2). However the shape of the light curve, derived mass ratio q $\sim$0.106 and f $\sim$ 72\% indicate that the variable is a low mass ratio overcontact binary with a high degree of geometrical contact, a unique characteristic of A-type W UMa systems. The best fit solutions indicate the presence of a dark spot on primary or secondary (Table 2). The presence of spots do not affect the mass ratio of a system and hence an A-type classification is justifiable. In general, the occultation at secondary minimum suggests that the primary is hotter than the secondary component but our solutions indicate that the secondary is slightly hotter. This phenomenon is probably caused due to the migration of the spot on the stellar surface, responsible for causing variable O'Connell effect and could {\bf account} for reversed minima, making an A-type to a W-type and vice-versa. For eg. Qian \& Yang (2005) reported A-type to W-type and then W-type to A-type features in the light curves of the overcontact binary FG Hya. Overall, there are three such low mass ratio overcontact binaries which resembles these features viz. FG Hya (Zola et al. 2010; Qian \& Yang 2005), V802 Aql (Samec et al. 2004; Yang et al. 2008) and V902 Sgr (Samec \& Corbin 2002). If the dark spot/spots are responsible for the ambiguity of V1's classification (A-type or W-type), a radial velocity spectroscopy is necessary to unveil the true nature of the source. 
  
The combination of low mass ratio  (q $\le$ 0.25) and high filling factor in overcontact binaries (f $\ge$ 50 \%) are considered to be the progenitors for FK Com-type and blue straggler stars. The gradual decrease of mass ratio makes these systems evolve as single fast rotating stars due to the inducement of Darwin's instability when the orbital angular momentum is more than three times of spin angular momentum, i.e., J$_{orb}$ $\ge$ 3 J$_{spin}$ (Hut 1980). The merging is also possible when the degree of overcontact is high which helps in causing the dynamical instability (Rasio \& Shapiro 1995). Table 3 lists the high filling factor (f $>$ 50\%) and low mass ratio (q $<$ 0.25) overcontact binary systems. For these systems, we studied the correlation between the mass ratio and period (Fig. 5). Interestingly, we found systematic trends along the mass ratio with a cut-off. 
We fitted a broken power law model  $\frac {(x/p1)^p2} {(x/p1)^(p2 - p3)+1} \times (p4)/2$, where p1 is break (a value of log q), p2 and p3 are power-law indexes before and after the break and p4 is the normalization. We found that the break is at log q = -1.071 $\pm$ 0.04 and the error bar   indicates the 90 \% confidence limit. The line in Figure 5 shows the best fit and obtained a cut-off mass ratio q = 0.085. The mass transfer from  primary to secondary results in a short period high mass ratio systems (i.e. trend below q$_{critical}$ $\le$ 0.085).  However above q$_{critical}$ $>$ 0.085, as a system has to evolve from high to low AM state, the simultaneous processes of mass transfer from  primary to secondary and mass loss of secondary component and partial transferred mass from Lagrangian point L2 can drive the system towards short period low mass system configuration. Moreover as period decreases, the critical Roche lobes shall be shrinking, making f to increase and finally evolve as a single rapid rotator supported from the sample (Table 3) i.e. high f values are associated with low mass ratio systems.

Theoretically a critical mass ratio has been predicted because till now no overcontact binaries with low q $<$ 0.07 have been discovered. Rasio \& Shaprio (1995) constrained a minimum mass ratio of 0.09 and while considering the rotation of the secondary, the critical mass ratio, q$_{critical}$ lowers to 0.076 (Li \& Zhang 2006). Assuming convective and radiative envelope in both secondary and primary, the q$_{critical}$ was found to be around 0.094--0.109 (Arbutina 2007) and taking into account differential rotation of the primary, the q$_{critical}$ converges at 0.070--0.074. The mass ratio of the variable V1 is found to be $\sim$0.106, close to the theoretically predicted values. We have also plotted the period-colour relation for 110 overcontact binary systems from the data of Terrell et al. (2012). The line in Figure 6 shows the short period blue envelope (SPBE) theoretical line i.e. B-V$_\circ$ = 0.04 $\times$P$^{-2.25}$ (Rucinski 1997) and the variable V1 is found to be above it. Any overcontact binary above SPBE line has to further evolve towards longer period and cooler temperature (towards lower right in Figure 6). Overall, the variable V1 is an important overcontact binary source as it has the low mass ratio and high fill-out factor (see Table 3) and these kinds of sources are considered to be at a stage of merger, potentially forming a FK-Com/blue straggler type star. Moreover the solutions suggest that the a dark spot is affecting the occultation at secondary minimum and hence V1 is an interesting source for future observations which may unveil the variable O'Connell effect.

The H$\alpha$ 6563\AA\ line emission in overcontact binaries is sparsely studied. This line along with Ca H and K, Fe II and the Mg II resonance line in the UV are the primary indicators of chromospheric activity in solar type stars.The strong absorption of the H$\alpha$ line forms the basis of a zero point for the measure of chromospheric activity, however few stars are found to be associated with the weak level of activity and shows weak absorption.
The fill-in effect is generally explained as the consequence of chromospheric activity which decreases the equivalent width of the H$\alpha$ absorption line. This scenario was first observed in overcontact binaries V566 Oph, AH Vir and W UMa by Barden (1985). The H$\alpha$ variation in VW Cep was studied by Barden (1985), Frasca et al. (1996) and Kaszas et al. (1998) and concluded that activity is associated with the primary rather than secondary component in overcontact binaries.    
From Fig. 4 it is clear that at phase 0.5 the H$\alpha$ is more filled-up suggesting a high level of chromospheric activity associated with the primary. The conclusion was based on the visual inspection, hence a continues spectroscopic monitoring is needed to explain the nature of the H$\alpha$ profile.


Because H$\alpha$ line EW varies due to chromospheric activity, we studied the correlation between period and EW for a few overcontact binaries (Fig. 4). It is clear that as the orbital period decreases (i.e fast rotation), the EW of the H$\alpha$ line is found to be decreasing (which is measure of increasing activity), leading  towards the saturation at lower periods. 

\subsection{Distance and Absolute parameters}
Overcontact binaries are potential sources to estimate the distance of binary systems. Rucinski \& Duerbeck (1997) introduced {\bf a} calibration scheme to determine the distance of overcontact binaries and the best calibration is given by  
M$_{V}$= -4.44 Log P + 3.02 (B-V)$_{o}$ + 0.12, where M$_{V}$ can be predicted with an uncertainty of up to $\pm$ 0.25 mag.
Using this relation and m$_{V}$ = 11.61, the distance was estimated to be $\sim$279 pc. Gettel et al. (2006) also estimated the distance to the variable to be $\sim$306 pc based on the relation Log D = 0.2 V$_{max}$-0.18 log (P) -1.60 (J-H) + 0.56. Considering the error of $\pm$0.25 mag in the value of M$_{V}$, the difference in estimated distances is found to be within the error bars.

 Based on well studied overcontact binaries, Gazeas (2009) obtained three dimensional correlations as stated below:\\
 log $M_{1}$ = 0.725 log P - 0.076 log q + 0.365 \\
 log $M_{2}$ = 0.725 log P + 0.924 log q + 0.365 \\
 log $R_{1}$ = 0.930 log P - 0.141 log q + 0.434 \\
 log $R_{2}$ = 0.930 log P + 0.287 log q + 0.434 \\
 log $L_{1}$ = 2.531 log P - 0.512 log q + 1.102 \\
 log $L_{2}$ = 2.531 log P + 0.352 log q + 1.102 \\

Using these correlations (Gazeas 2009), the absolute parameters obtained are $M_1$= 1.10 $\pm$ 0.08 M$_{\odot}$, $M_2$= 0.17 $\pm$ 0.05 M$_{\odot}$, $R_1$= 1.15 $\pm$ 0.04 R$_{\odot}$, $R_2$= 0.42 $\pm$ 0.02 R$_{\odot}$, $L_1$= 1.66 $\pm$ 0.24 L$_{\odot}$ and $L_2$= 0.22 $\pm$ 0.02 L$_{\odot}$.

In conclusion, we report the photometric solution and variation of the H$\alpha$ line of a overcontact binary ASAS J082243+1927.0. The high value of f and low q suggests that it is a low mass ratio overcontact binary with high degree of contact, making it as one of the potential sources for merger studies. We found a cut-off mass ratio in mass ratio--period plane at q$_{critical}$ = 0.085, closely matching with various theoretically predicted ones. The H$\alpha$ line varied at different phases showing the relatively fill-in effect at secondary minimum and is arising arguably from the primary component. A small sample of H$\alpha$ EW and period of overcontact binaries shows a correlation indicating a saturation towards chromospheric emission. A spectroscopic radial velocity study is essential to resolve the ambiguity of the variable's classification (A-type or W-type), to confirm the photometric mass ratio and to constrain the explicit nature of filled-in H$\alpha$ line. Moreover a long term photometric study of this variable is necessary to know its secular period variation and explore the variable O'Connell effect.

\section{Acknowledgements}
 We acknowledge the Director of IUCAA, Pune and the Director of IIA, Bengaluru for allocating the time for the observations at IGO and VBO. 
We also thank Dr. Vijay Mohan, IUCAA for helping us with the observations. We thank the referee for the valuable comments and suggestions which improved the quality of the paper. K.S acknowledge the support from UGC-BSR Research Start-up grant, Government of India .

\clearpage

\begin{figure}
\includegraphics[height=12cm,width=18cm,angle=0]{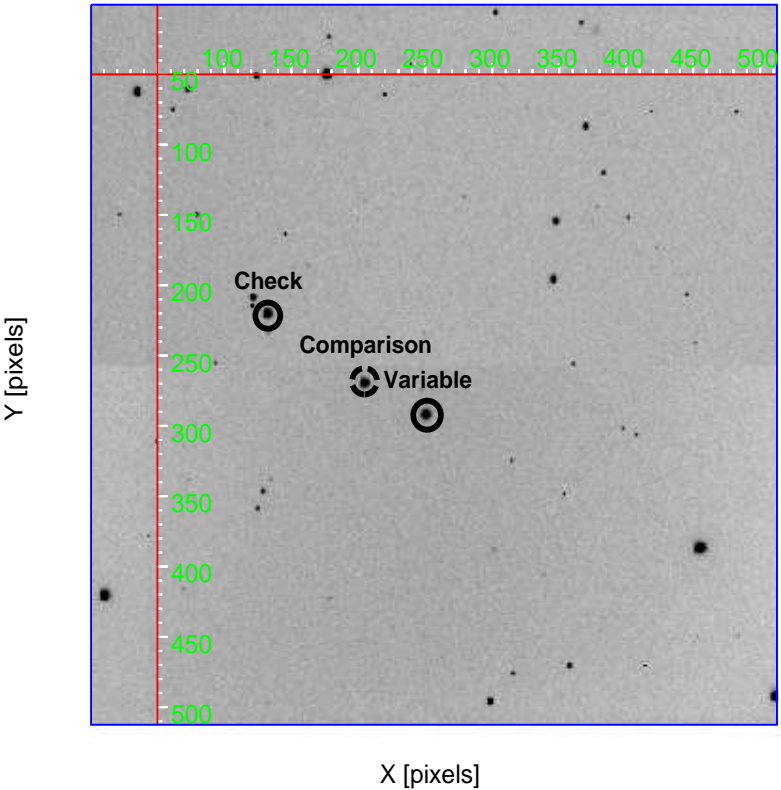}
     \caption{The field of the variable ASAS J082243+1927.0. North is up and East is to left. } 
 \end{figure}
\clearpage
\begin{figure}
\includegraphics[height=12cm,width=18cm,angle=0]{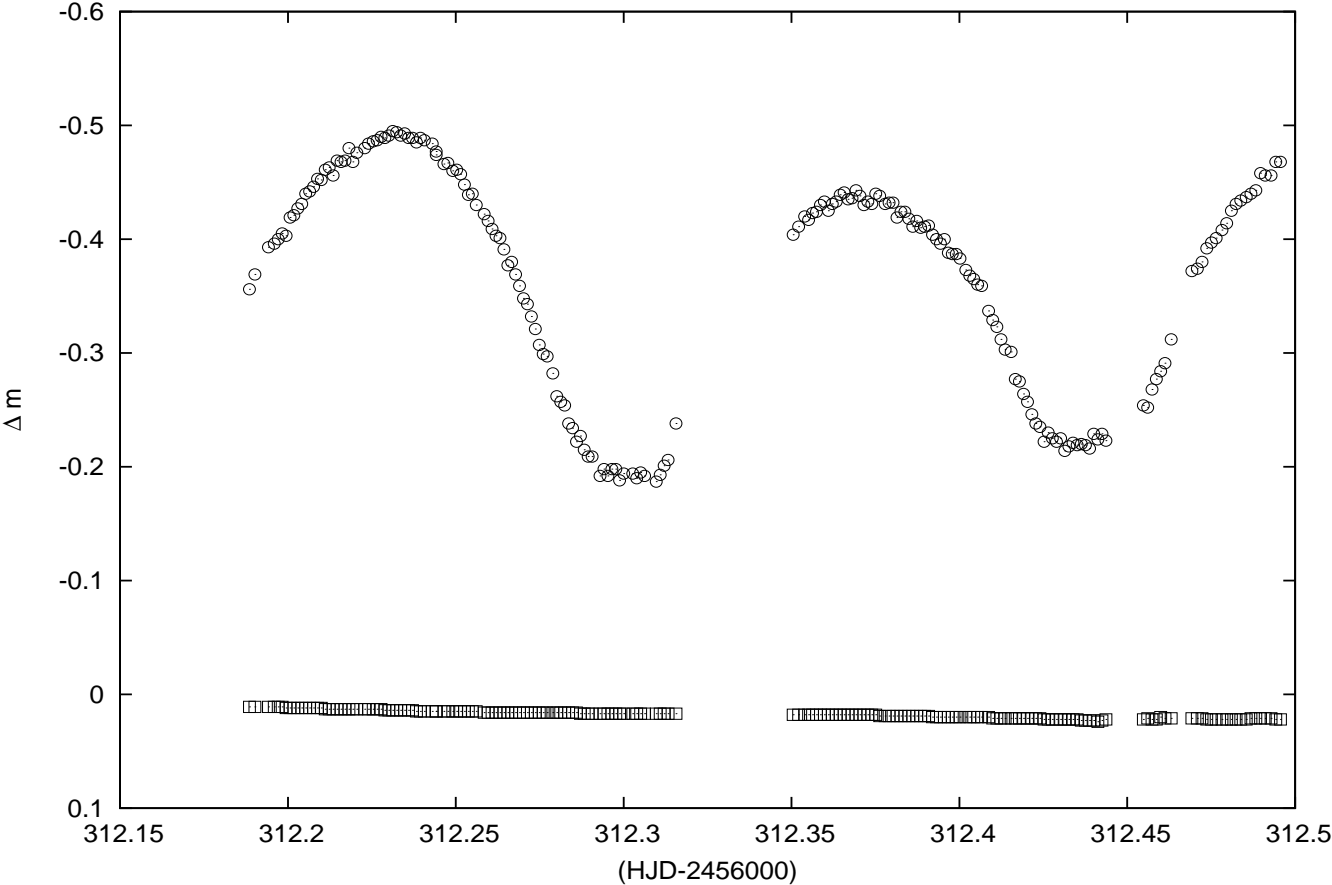}
     \caption{Magnitude difference (Bessell's V band) of the variable - comparison (circles) and check - comparison (squares) observed on 19 January, 2013 .} 
 \end{figure}

\clearpage

\begin{figure}
\includegraphics[height=13cm,width=7.5cm,angle=-90]{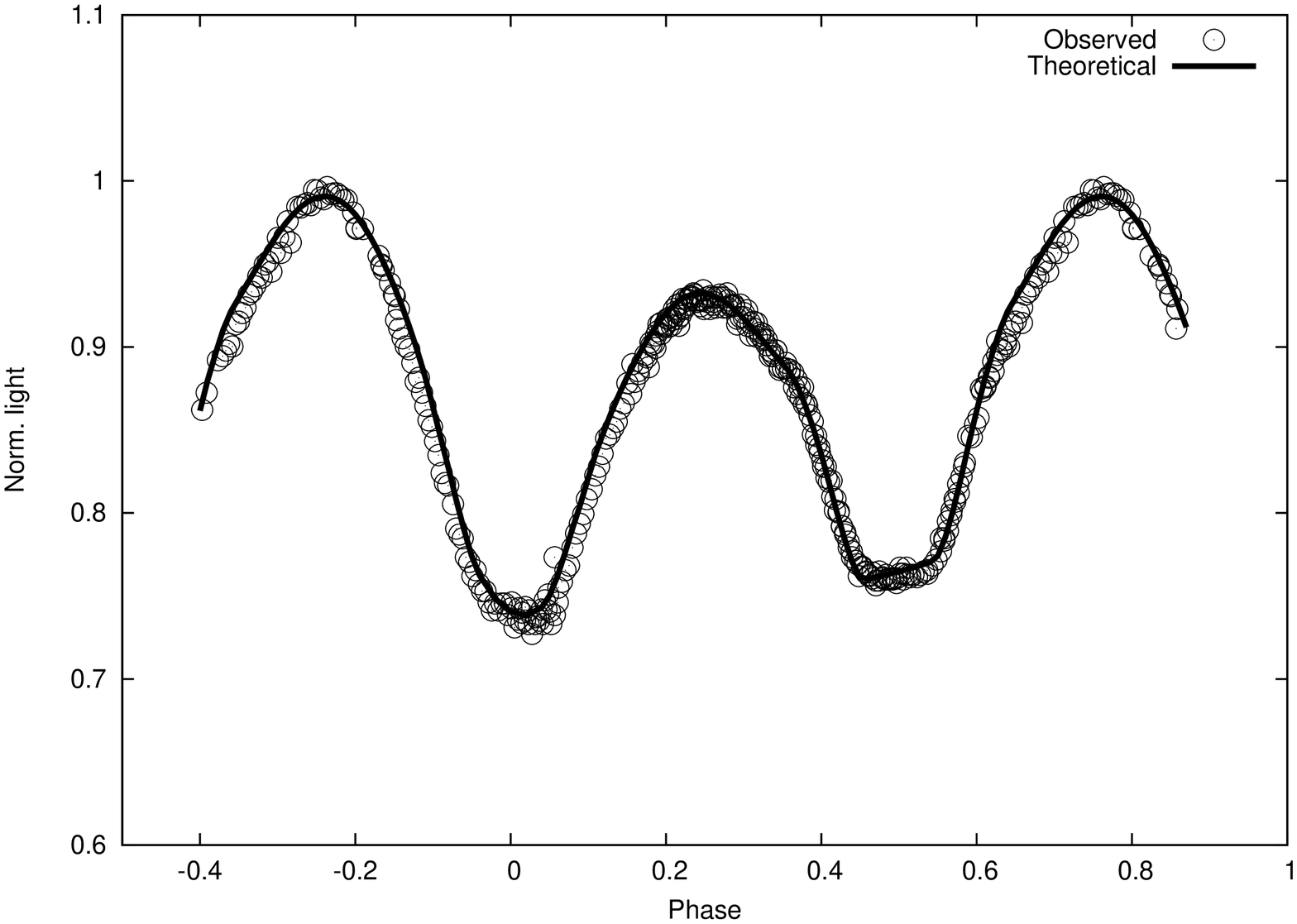}
\includegraphics[height=13cm,width=7.5cm,angle=-90]{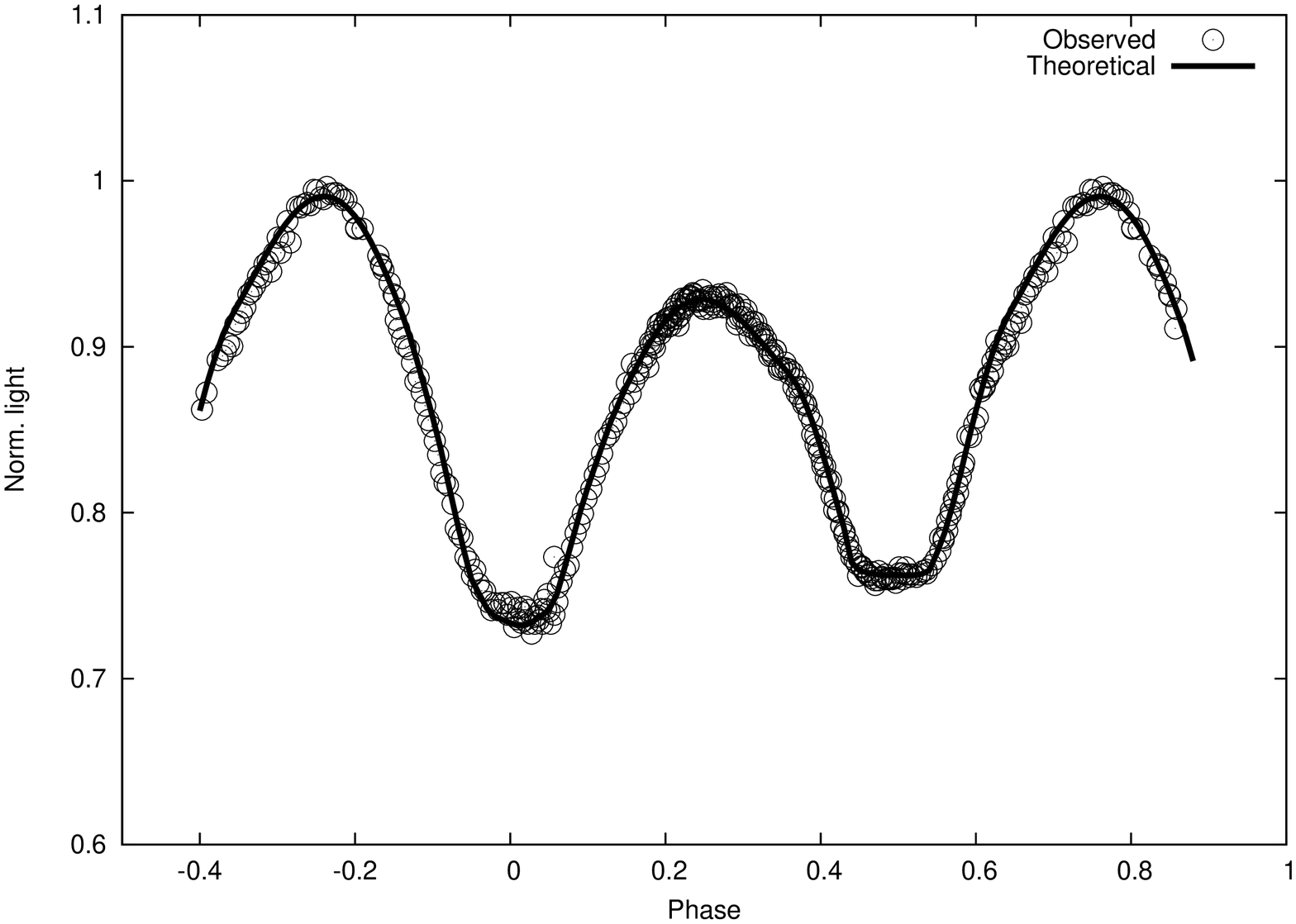}

     \caption{Top: The V band light curve and theoretical fit using an unspotted solution and a dark spot over primary. Bottom: Similar figure with the same unspotted solution and a dark spot over secondary.} 
       \label{Fig3}
 \end{figure}
\clearpage
\begin{figure}
\includegraphics[height=18cm,width=15cm,angle=-90]{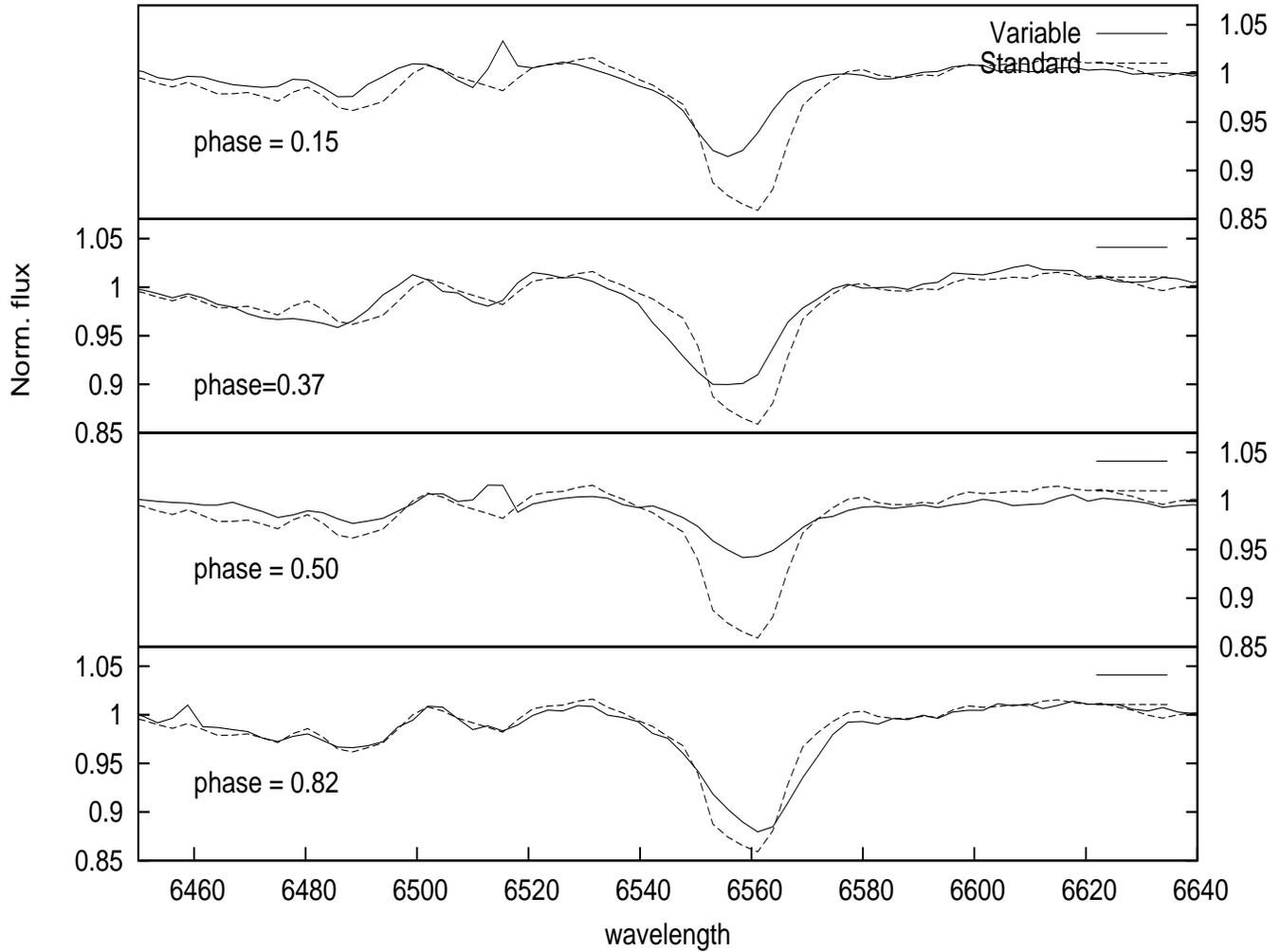}\\
\includegraphics[height=18cm,width=6cm,angle=-90]{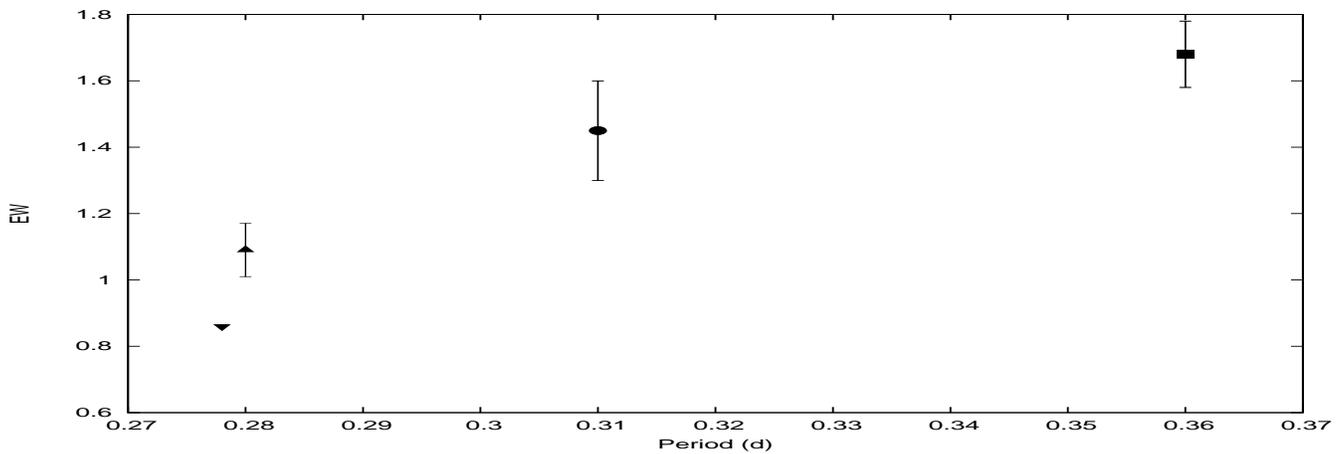}\\
     \caption{Top panel: Comparison of variable's H$\alpha$ line with standard star at various phases. 
 Bottom panel: Period vs intrinsic equivalent width of overcontact binaries (see text). Variable V1 is represented as a triangle.} 
        \end{figure}
\clearpage

\begin{figure}
\includegraphics[height=18cm,width=20cm,angle=-90]{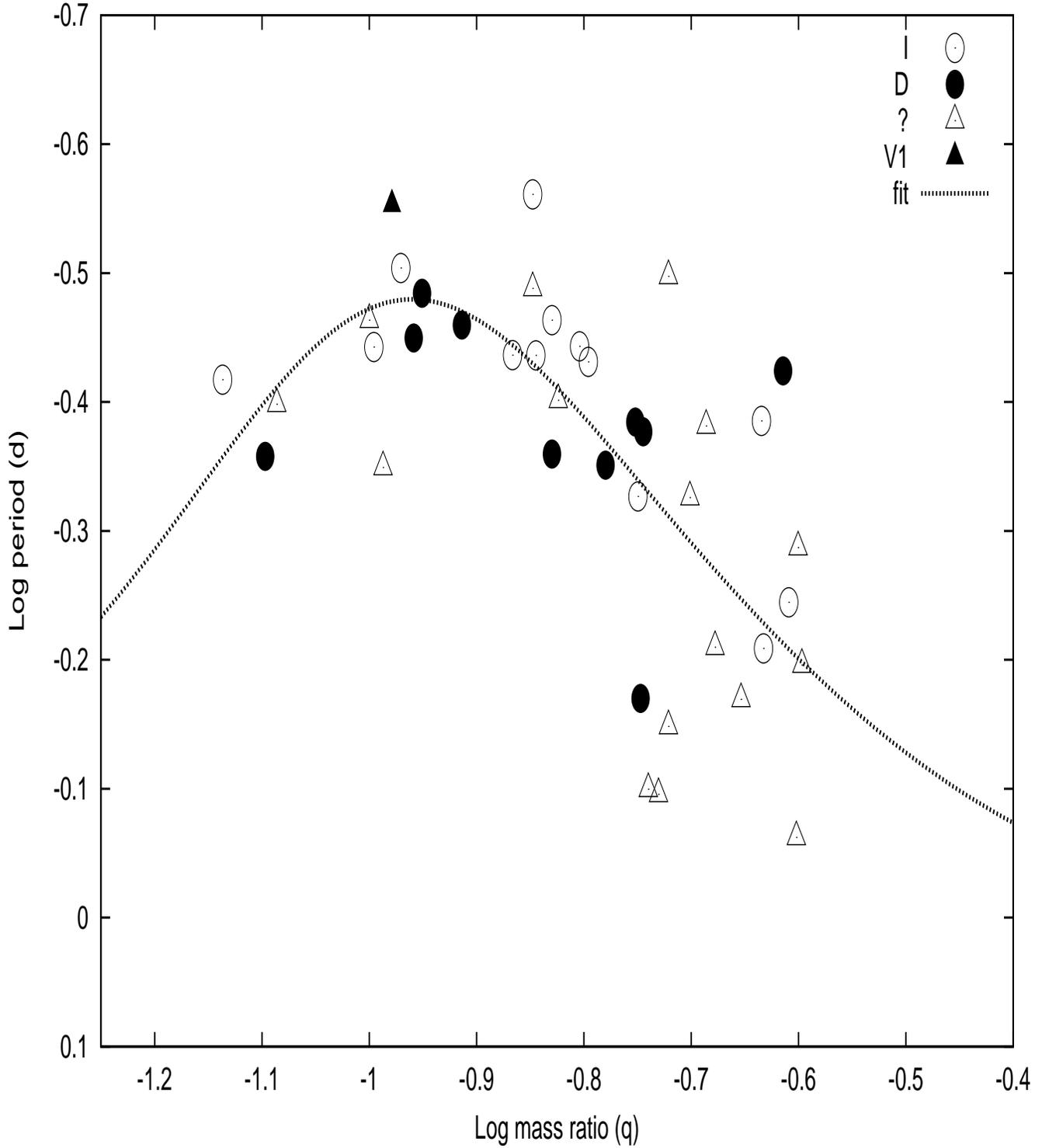}
     \caption{Mass ratio vs period for high filling factor, low mass ratio overcontact binaries. Open and filled circles denote the secular period increase (I) and decrease (D) for the individual systems (see Table 3). The open triangle (?) denotes the systems where no secular period variation were reported or observed and filled triangle shows the position of V1. Dashed line shows the best fit broken power law model (see text).} 
       \label{Fig4}
\end{figure}
\clearpage
\begin{figure}
\includegraphics[height=18cm,width=20cm,angle=-90]{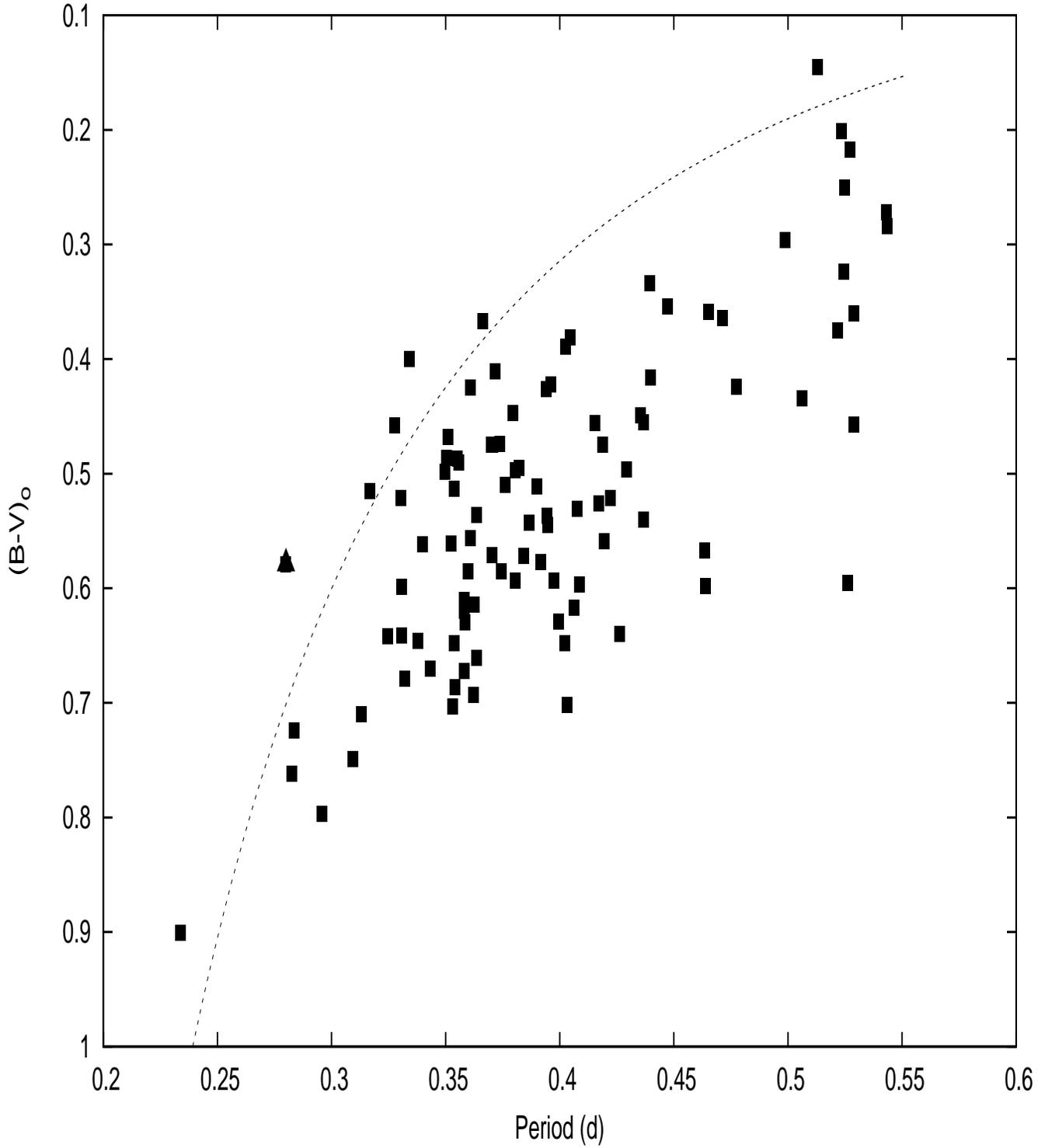}
     \caption{Period vs intrinsic colour for overcontact binaries and the line shows the theoretical SPBE. The triangle represents V1.} 
\end{figure}
\clearpage
{\begin{table*}
\begin{minipage}[t]{\columnwidth}
\caption[solutions]{Details of comparison and check stars}
\label{Table 1}
\renewcommand{\footnoterule}{}
\centering
\begin{tabular}{ccccccc}
\hline
\hline
Name &V&  B-V \\ \\
\hline
\hline
V1 (Variable)&11.62 $\pm$0.13&0.57$\pm$ 0.01\\
TYC 1386-1630-1 (Comparison) & 11.58 $\pm$0.13&0.48 $\pm$0.08\\
TYC 1386-121-1 (Check) & 11.99 $\pm$ 0.21 &0.54 $\pm$0.08 \\   
\hline
\hline
\end{tabular} 
\end{minipage}
\end{table*}}

{\begin{table*}
\begin{minipage}[t]{\columnwidth}
\caption[solutions]{The photometric solutions obtained for V1 using W-D method.}
\label{Table 1}
\renewcommand{\footnoterule}{}
\centering
\begin{tabular}{ccccccc}
\hline
\hline
Parameters & &  unspotted &  Dark 1 & Dark 2 \\ \\
\hline
\hline
A$_{1}$ = A$_{2}$& & 0.50 &  0.50 &0.50\\ 
g$_{1}$ = g$_{2}$& & 0.32 &  0.32 &0.32\\ 
$T_{1}$\,(K)& & 5960 &  5960&  5960\\
$T_{2}$\, (K)& & 6080 $\pm$34 & 6078 $\pm$ 30 &6038$\pm$28\\
q & & 0.110 $\pm$0.004&  0.106$\pm$0.002&0.105$\pm$0.002\\ 
i$^{o}$& & 75.60 $\pm$1.04& 76.58$\pm$1.04&75.58$\pm$1.05\\ 
$\Omega_{1,2}$ & &1.9218 $\pm$0.0039& 1.9238$\pm$0.0036&1.9320$\pm$0.0084\\

fill-out factor (f \%)& & 73.82 $\pm$1.32& 71.91$\pm$1.13&61.28$\pm$1.89\\
$r_{1}$ & pole & 0.5278 $\pm$0.0036& 0.5442$\pm$0.0012   &0.5357$\pm$0.0016\\
        & side & 0.5865  $\pm$0.0055& 0.6128 $\pm$0.0019 &0.5989$\pm$0.0025\\
        & back & 0.6068 $\pm$0.0057&  0.6349 $\pm$0.0020 &0.6199$\pm$0.0024\\

$r_{2}$ & pole & 0.1985 $\pm$0.0154 & 0.1799$\pm$0.0054 &0.1879 $\pm$0.0090\\
 	& side & 0.2062 $\pm$0.0182 & 0.1856$\pm$0.0062 &0.2059 $\pm$0.0109\\
	& back & 0.2365 $\pm$0.0348&  0.2062$\pm$0.0101 &0.2089 $\pm$0.0219\\

$L_1/L_1 + L_2$& &0.8773$\pm$0.0038&0.8738$\pm$0.0024&0.8739$\pm$0.0036\\

Spot Colatitude ($^\circ$)& & &56.33$\pm$1.24 &  79.45 $\pm$1.37 \\
Spot Longitude ($^\circ$)&&&284.81$\pm$4.23&118.99$\pm$4.82\\
Spot Radius($^\circ$)&&&20.28$\pm$0.52&30.93 $\pm$1.69\\
T$_{Spot}$/T$_{local}$  &&&  0.74 $\pm$0.04 &0.64 $\pm$0.07\\


$\Sigma$w(o-c)$^{2}$ &&0.0088  &0.0024&0.0022 \\
 
\hline
\hline
\end{tabular}

\end{minipage}
\end{table*}}


{\begin{table*}
\begin{minipage}[t]{\columnwidth}
\caption[solutions]{List of high filling factor low mass ratio overcontact binary systems. 
I represents the secular period increase, D secular decrease and "?" no secular variations.}
\label{Table 3}
\renewcommand{\footnoterule}{}
\centering
\begin{tabular}{cccccccc}
\hline
\hline
Name & mass ratio (q) &  degree of overcontact (f \%) &  period (d) (dp/dt) &  References\\
\hline
\hline
J13031-0101.9          &  0.150 & 50.0  &  0.2709  (?) & Pribulla et al. (2009)\\
V345 Gem	&	  0.142 & 72.0  &  0.2747 (I) & Yang et al. (2009)\\
{\it ASAS 082243+1927.0 (V1)}&  0.106 & 72.0  & 0.2800 (?) & Present study\\
V119 Cygni               & 0.107 & 68.6  &  0.3133 (I) & Ulas et al. (2012)\\
ASAS 021209+2708.3&       0.19  & 58.7  &  0.3181 (?)& Acerbi et al. (2011)\\
V677 Cen	&	  0.142 & 57.0  &  0.3250 (?)&  Kilmartin et al. (1987)\\
FG Hya          &         0.112 & 85.6  &  0.3278 (D)&Qian \& Yang (2005)\\
EM Psc          &         0.148 & 95.0  &  0.3439 (I)&Qian et al. (2008)\\
GSC 619-232     &         0.100 & 93.4  &  0.3439 (?) & Yang et al. (2005)\\
GR Vir          &         0.122 & 78.6  &  0.3469  (D) &Qian \& Yang (2004)\\
CK Boo		&	  0.110 & 91.0  &  0.3551 (D) & Nelson et al. (2014)\\
AH Cnc          &          0.157&  73.0 &   0.3604 (I)& Qian et al. (2006)\\
AW CrB          &          0.101&75   &     0.3609 (I) &Broens (2013)\\ 
DZ Psc          &          0.145&  79.0 &   0.3661 (I)&Yang et al. (2013)\\
V410 Aur        &          0.143&  60.0 &   0.3663 (I) & Qian et al. (2005)\\
XY Boo          &          0.160&  50.0 &   0.3705 (I)& Qian et al. (2005)\\
YY CrB          &          0.243 & 63.4 &   0.3765 (D) & Essam et al. (2010)\\
V857 Her        &          0.073 & 80.0 &   0.3825 (I) & Qian et al. (2005)\\
TYC 4157-0683-1 &          0.150 & 76.3  &0.3969 (?)  &Acerbi et al. (2014)\\
V870 Arae       &          0.082 &  96    &0.3997 (I) & Szalai et al. (2007) \\
QX Andromedae	&          0.232 & 56   &   0.4117 (I) &Qian et al. (2007)\\
CU Tau          &          0.177 & 50.1 &   0.4125 (D) &Qian et al. (2005)\\
TYC 3836-0854-1 &          0.206    &59.2  &   0.4155 (?) &Acerbi et al. (2014)\\
Y Sex           &          0.180 & 64.0 &   0.4198 (D) & Qian et al. (2000)\\
AW UMa          &          0.080 & 84.6 &   0.4387 (D) & Pribulla et al. (1999)\\
XY LMi   	&  	  0.148  &74.1  &  0.4368  (D) & Qian et al. (2011)\\
TV Mus         &           0.166 & 74.3&    0.4456 (D) & Qian et al. (2005)\\
DN Boo 		&	   0.103&   64.0 & 0.4475 (?)& Senavci et al. (2008)\\
V728 Her       &           0.178 & 71.4&    0.4712 (I) &Qian (2001b)\\
V409 Hya       &           0.199 & 67.0&    0.4722 (?) & Labadorf et al. (2009)\\
AH Aur         &           0.169 & 62.5&    0.4941 (?) &Rucinski \& Lu (1999)\\
BU Vel         &           0.251 & 61.0&    0.5162 (?) & Qian (2001a)\\
AP Aur         &           0.246 & 64.6&    0.5693 (I) & Li et al. (2001)\\
DN Aur         &           0.210 & 54.0&    0.6168 & Goderya et al. (1997a)\\
V 1853 Ori     &           0.203 & 54.0   &   0.6168 & Semac et al. (2011)\\
UZ Leo         &           0.233 & 84.8 &   0.6180 (I) & Qian et al. (2001)\\
IK Per	       & 	  0.179  &63.0  &  0.6760 (D) & Zhu et al. (2005)\\
FN Cam         &           0.222 & 88.0 &   0.6771 (?) &  Pribulla \& Vanko (2002)\\
GSC 1720-658   &           0.253 & 65.8 &   0.6368 (?) &  Maciejewski et al. (2003)\\
HV UMa         &           0.190 & 77.0 &   0.7107 (?) &  Csak et al. (2000)\\
MW Pav         &           0.182 & 50.4 &   0.7949 (?) &  Lapasset (1980)\\
V2388 Oph      &           0.186 & 65.0  &  0.8023  (?) & Yakut et al. (2004)\\
KN Per          &          0.250  &54.5   & 0.8664 (?) &  Goderya et al. (1997b)\\

\hline
\hline
\end{tabular} 
\end{minipage}
\end{table*}}

\clearpage

\label{lastpage}

\end{document}